\documentstyle[preprint,aps]{revtex}

\begin{document}

\preprint{FIMAT-1/95}

\draft

\title{X-ray reflectivity of Fibonacci multilayers}

\author{F.\ Dom\'{\i}nguez-Adame and E.\ Maci\'a \thanks{Also at the
Instituto de Estudios Interdisciplinares, El Guijo, Z4 Galapagar,
E-28260 Madrid, Spain.}}

\address{Departamento de F\'{\i}sica de Materiales,
Facultad de F\'{\i}sicas, Universidad Complutense,
E-28040 Madrid, Spain}

\date{\today}

\maketitle

\begin{abstract}

We have numerically computed the reflectivity of X-ray incident normally
onto Fibonacci multilayers, and compared the results with those obtained
in periodic approximant multilayers.  The constituent layers are of low
and high refractive indices with the same thickness.  Whereas
reflectivity of periodic approximant multilayers changes only slightly
with increasing the number of layers, Fibonacci multilayers present a
completely different behaviour.  In particular, we have found a
highly-fragmented and self-similar reflectivity pattern in Fibonacci
systems.  The behaviour of the fragmentation pattern on increasing the
number of layers is quantitatively described using multifractal
techniques.  The paper ends with a brief discussion on possible
practical applications of our results in the design of new X-ray
devices.

\end{abstract}


\narrowtext

\section{Introduction}

The realization of well-controlled quasiperiodic superlattices
\cite{Merlin,Todd} has led to a widespread theoretical interest in the
study of one-dimensional quasiperiodic systems
\cite{Cohn,Angel,Chakrabarti}.  From the very beginning, most
researchers have considered the Fibonacci sequence as the archetypal of
a quasiperiodic system \cite{Koh,Ostlund}.  This point of view is firmly
established by X-ray diffraction analyses, which clearly reveal the
quasiperiodic nature of Fibonacci superlattices, even if moderately
large growth fluctuations during sequential deposition are present
\cite{Merlin,Cohn}.  It is by now well known that Fibonacci lattices
exhibit highly-fragmented energy spectra with a hierarchy of splitting
subbands displaying self-similar patterns \cite{Laruelle,Enrique,PLA}.
These novel features are directly related to the peculiar topological
order displayed by the underlying structure, namely its quasiperiodic
order \cite{PRE}.  But it is most important to stress that new and
striking phenomena are not only found in the case of electron dynamics.
In fact, harmonic vibrations in quasiperiodic lattices also present
highly-fragmented and self-similar frequency spectra \cite{phonon}.  In
addition, theoretical studies on incoherent \cite{incoherent} and
Frenkel \cite{Frenkel} excitons in Fibonacci systems have revealed a
very different dynamics in comparison to that shown in both random and
periodic lattices.

In this letter we consider the X-ray reflectivity of a multilayered
system where the refractive indices of the layers are arranged according
to the Fibonacci sequence.  The aim of this paper is twofold.  In the
first place, we carry out a theoretical investigation of Fibonaccian
systems from a different perspective to shed more light onto the role
played by quasiperiodic order in their physical properties.  In the
second place, our present study suggests new practical applications of
Fibonacci systems.  In particular, we shall demonstrate that Fibonacci
multilayers could be used as selective filters of soft X-ray, allowing
for a fine tuning of different narrow lines.  This may be compared with
recent results reported on multilayers with randomly varying
thicknesses, which have been proposed as a broad bandwidth X-ray mirror
\cite{Yoo}

The system we study in this work is a Fibonacci multilayer (FM)
consisting of two different kinds of layers of the same thickness $d$.
Layers A (B) have low (high) refractive indices $n_A$ ($n_B$).  For the
sake of simplicity we neglect the absorption of X-ray incident normally
onto the sample surface, so that the refractive indices are real
parameters.  Finally, we assume that the whole system is placed in
vacuum.  In general, a Fibonacci system of order $l$ is generated from
two basic units A and B by successive applications of the inflation rule
A~$\rightarrow$~AB and B~$\rightarrow$~A.  This sequence comprises
$F_{l-1}$ elements A and $F_{l-2}$ elements B, $F_l$ being the {\em
l\/}th Fibonacci number given by the recurrent law $F_l=F_{l-1}+F_{l-2}$
with the initial values $F_0=F_1=1$.  As $l$ increases the ratio
$F_{l-2}/F_{l-1}$ converges toward the so called inverse golden mean
$\tau=(\sqrt{5}-1)/2\sim 0.618\ldots$\ The reflectivity of the FM
structure is then easily computed numerically using the Rouard's method
(see, e.g., Ref.~\onlinecite{Yoo}).  To facilitate direct comparison
with previous studies of Yoo and Cue in random multilayers \cite{Yoo},
we have taken the same physical parameters, namely $d=50\,$\AA,
$n_A=0.9200$ and $n_B=0.9995$.  In this way we can separate those
features of the reflectivity spectrum stemming from the underlying
quasiperiodic order in a straightforward manner.

In periodic multilayers arranged according to the refractive indices
sequence $n_An_Bn_An_B\ldots$, the reflectivity shows a pronounced peak
centered at $184\,$\AA, with a bandwidth of about $11\,$\AA \cite{Yoo}.
Note that in this case the ratio $c$ between the number of high and low
refractive index layers is $c=1/2$.  This value is not close to $\tau$,
so that it is difficult to carry out a direct comparison with the
reflectivity obtained in FMs. Instead, it becomes more appropriate to
consider periodic approximants to the quasiperiodic FM. To this end we
have constructed periodic structures by a repetition of blocks of the
form ABAAB ($c=0.667$), or ABAABABA ($c=0.600$), or ABAABABAABAAB
($c=0.625$), namely the $4\,$th, $5\,$th and $6\,$th order approximants
to the FM. The number of blocks is repeated in any realization until the
total number of layers $N$ roughly equals the desired Fibonacci number.
Figure~\ref{fig1} shows the reflectivity $R(\lambda)$ corresponding to
these periodic approximants (panels a, b and c) and to the quasiperiodic
$N=F_{12}=233$ FM (panel d).  In all cases, two major reflection peaks
are clearly observed at about $\lambda=150\,$\AA\ and
$\lambda=250\,$\AA, whose bandwidths are $4\,$\AA\ and $11\,$\AA\
respectively, and a less pronounced peak close to $\lambda=125\,$\AA.
The position at which these major peaks are centered varies slightly
depending on the order of the approximant considered.  In addition to
those major peaks, a set of {\em subsidiary} peaks displaying high
reflectivity arises in periodic approximants, the number of them being
increasingly large as the order of the approximant increases and,
consequently, the envelope of $R(\lambda)$ is less uniform.

The origin of these subsidiary peaks can be easily visualized from a
closer inspection if Fig.~\ref{fig1}.  Let us start by noting that no
subsidiary peaks appear between the two major peaks in the reflectivity
pattern corresponding to the $4\,$th order approximant (panel a).  As
soon as we increase by one step the order of the approximant considered,
{\em one} prominent subsidiary peak arises between them (panel b).  By
increasing two steps the order of the approximant, {\em two} subsidiary
peaks appear instead meanwhile {\em one} minor subsidiary peak develops
between the major peaks centered at about $\lambda=150\,$\AA\ and
$\lambda=250\,$\AA\ (panel c).  By further increasing the order of the
approximant, an increasing number of subsidiary peaks progressively
appear (this number is $F_{n-3}-1$, $n\geq 3$ being the order of the
considered approximant), until we obtain the reflectivity pattern of the
quasiperiodic FM (panel d).  The resulting pattern forms a dense set of
sharp and narrow peaks, some of them presenting reflectivity larger than
$50\%$.

Self-similarity and multifractal properties are both characteristic
features of quasiperiodic orderings.  Then it follows in a natural way
to look for such properties in the reflectivity of FMs. In
Fig.~\ref{fig2} we compare the reflectivity pattern of a FM containing
$F_{10}=89$ layers with the central portion of the reflectivity pattern
of a FM containing $F_{13}=377$ layers.  This figure clearly shows the
self-similar characteristics of the reflectivity peaks, that is to say,
a given interval of wavelengths of a short FM is mapped onto a small
interval of a larger FM. The rescaling procedure relates FMs with
$F_{l}$ and $F_{l+3}$ layers, in an analogous fashion to self-similar
electronic spectra \cite{Kohmoto}.

In view of this result, we believe that it is important to get a
quantitative estimation of the fragmentation as $N$ increases.  To be
specific, it is clear that reflectivity of periodic approximants of any
order changes very little on increasing the number of layers, whereas a
hierarchical fragmentation process takes place in FMs. This is similar,
although not indentical, to what is found in the case of electronic
properties (energy spectra and wave functions) in Fibonacci lattices.
Following this analogy, we make use of multifractal analysis to get
insight into the fragmentation of the reflectivity pattern.  In
particular, the participation ratio as defined, for instance, in
Ref.~\cite{Mato}, has been successfully used to describe the spatial
nature (extended or localized) of electron wave functions.  This method
is readily generalized to the case of any positive measure defined in
the system.  Particularly, we can apply the same concepts to
$R(\lambda)$ since is a positive-defined quantity.  Thus we introduce
the participation ratio for the reflectivity as $ P(N) = \left[ \int\>
R(\lambda)\,d\lambda \right]^2/\left[ \int\>
R^2(\lambda)\,d\lambda\right]$.  The value of $P(N)$ gives an estimation
of the overall reflectivity of the sample as a function of the number of
layers: The higher its value the higher the whole reflectivity.  It is
worth mentioning that performing the numerical integration requires very
tiny integration steps since $R(\lambda)$ presents a more and more
detailed structure on increasing $N$ due to the hierarchical
fragmentation scheme previously discussed.  Thus one must repeat
numerical integration with smaller and smaller integration steps until
convergence is reached.  Typically, $1.6\times 10^4$ grid points are
needed for a maximum number of layers of $F_{18}=4181$.
Figure~\ref{fig3} shows the results obtained for both the $4\,$th order
approximant and the quasiperiodic FM as a function of the number of
layers.  Notice that periodic multilayers present an almost constant
value of $P$ for different number of layers, in agreement with the fact
that the envelope of $R(\lambda)$ remains almost unchanged on increasing
$N$ in those systems.  On the contrary, the value of $P$ in FMs
increases monotonously with $N$ and, in addition, it is always larger
than the corresponding value for periodic approximants of the same size.
Therefore, we can conclude that the characteristic fragmentation process
observed in quasiperiodic multilayers leads to an overall increase of
their X-ray reflectivity.  This interesting property can be directly
related to the self-similarity displayed by the reflectivity pattern.
In fact, both the total number and the average height of subsidiary
peaks progressively increase and new peaks appear after inflating the
multilayer structure.

To summarize, we have numerically evaluated X-ray reflectivity of
Fibonacci multilayers and compared it with that corresponding to
periodic approximants.  The main result is that both systems present a
very different X-ray reflectivity incident normally onto the surface.
Whereas reflectivity of periodic multilayers changes only slightly on
increasing the number of layers, quasiperiodic multilayers present a
completely different trend.  We have observed a highly-fragmented
reflectivity pattern as a function of the incident wavelength.  Whenever
the order $l$ of the FM increases (i.e., the inflation rule is applied),
the subsidiary peaks increase their heights and new peaks arise.  These
new peaks also increase reflectivity on further increasing the system
size.  These conclusions has been established quantitatively by means of
multifractal analysis and, in particular, using the participation ratio
$P$.  The value of $P$ in periodic approximants remains constant on
increasing $N$ whereas in quasiperiodic systems increases monotonously.
Moreover, $P$ is always higher in the later systems.  Hence the overall
reflectivity is larger in this case.  Finally, some comments in regards
to multilayers with random thicknesses are in order.  Yoo and Cue have
recently demonstrated that random multilayers present a broad
reflectivity peak whose bandwidth broadens as fluctuations are stronger
\cite{Yoo}.  Hence, random multilayers increase the whole reflectivity
on increasing fluctuations, and the system acts as a broad bandwidth
mirror.  We have found that Fibonacci multilayers also behaves as a
mirror but, unlike random systems, they should be regarded as X-ray
selective filters instead.  Notice that the position of subsidiary peaks
could be changed by varying the refractive indices of the two
constituent layers and/or the corresponding thicknesses.  Therefore,
Fibonacci multilayers open new possibilities in the {\em engineering} of
soft X-ray devices.

This work has been supported by UCM under project PR161/93-4811.

\begin{figure}
\caption{X-ray reflectivity of $N$-multilayer structures with two basic
layers of refractive indices $n_A=0.9200$ and $n_B=0.9995$, each layer
thickness being $d=50\,$\AA.  Results for periodic approximants of (a)
$4\,$th order with $N=235$, (b) $5\,$th order with $N=232$, (c) $6\,$th
order with $N=236$, and (d) quasiperiodic FM with $N=F_{12}=233$ are
shown.}
\label{fig1}
\end{figure}

\begin{figure}
\caption{Self-similarity of X-ray reflectivity of quasiperiodic FM with
the same physical paramters $n_A$, $n_B$ and $d$ as in
Fig.~\protect{\ref{fig1}}. The number of layers are (a) $N=F_{10}=89$
and (b) $N=F_{13}=377$.}
\label{fig2}
\end{figure}

\begin{figure}
\caption{Participation ratio $P$ in units of $d$ as a function of the
number of layers $N$ in periodic approximants (dashed line) and
quasiperiodic FMs (solid line).}
\label{fig3}
\end{figure}


\begin{references}

\bibitem{Merlin} R.\ Merlin, K.\ Bajema, R.\ Clarke, F.\ -Y. Juang, and
P.\ K.\ Bhatacharya, Phys.\ Rev.\ Lett.\ {\bf 55}, 1768 (1985).

\bibitem{Todd} J.\ Todd, R.\ Merlin, R.\ Clarke, K.\ M.\ Mohanty, and
J.\ D.\ Axe, Phys.\ Rev.\ Lett.\ {\bf 57}, 1157 (1986).

\bibitem{Cohn} J.\ L.\ Cohn, J.\ J.\ Lin, F.\ J.\ Lamelas, H.\ He, R.\
Clarke, and C.\ Uher, Phys.\ Rev.\ B {\bf 38}, 2326 (1988).\

\bibitem{Angel} F.\ Dom\'{\i}nguez-Adame and A.\ S\'anchez, Phys.\
Lett.\ A {\bf 159}, 153 (1991).

\bibitem{Chakrabarti} A.\ Chakrabarti, S.\ N.\ Karmakar, and R.\ K.\
Moitra, Phys.\ Lett.\ A {\bf 168}, 301 (1992).

\bibitem{Koh} M.\ Kohmoto, L.\ P.\ Kadanoff, and C.\ Tang, Phys.\ Rev.\
Lett.\ {\bf 50}, 1870 (1983).

\bibitem{Ostlund} S.\ Ostlund and R.\ Pandit, Phys.\ Rev. B {\bf 29},
1394 (1984).

\bibitem{Laruelle} F.\ Laruelle and B.\ Etienne, Phys.\ Rev. B {\bf 37},
4816 (1988).

\bibitem{Enrique} E.\ Maci\'a, F.\ Dom\'{\i}nguez-Adame, and A.\
S\'anchez, Phys.\ Rev.\ B {\bf 49}, 9503 (1994).

\bibitem{PLA} F.\ Dom\'{\i}nguez-Adame, E.\ Maci\'a and B.\ M\'endez,
Phys.\ Lett.\ A {\bf 194}, 184 (1994).

\bibitem{PRE} E.\ Maci\'a, F.\ Dom\'{\i}nguez-Adame, and A.\ S\'anchez,
Phys.\ Rev.\ E {\bf 50}, 679 (1994).

\bibitem{phonon} M.\ Kohmoto and J.\ R.\ Banavar, Phys.\ Rev.\ B {\bf
34}, 563 (1986).

\bibitem{Naka} M.\ Nakayama, H.\ Kato, and S.\ Nakashima, Phys.\ Rev.\
B {\bf 36}, 3472 (1987).

\bibitem{incoherent} F.\ Dom\'{\i}nguez-Adame, E.\ Maci\'a, and A.\
S\'anchez, Phys.\ Rev.\ B {\bf 51}, 878 (1995).

\bibitem{Frenkel} E.\ Maci\'a and F.\ Dom\'{\i}nguez-Adame, Phys.\ Rev.\
B {\bf 50}, 16\,856 (1994).

\bibitem{Yoo} K.\ M.\ Yoo and N.\ Cue, Phys.\ Lett.\ A {\bf 195}, 271
(1994).

\bibitem{Kohmoto} M.\ Kohmoto, Phys.\ Rev.\ Lett.\ {\bf 51}, 1198
(1983).

\bibitem{Mato} G.\ Mato and A.\ Caro, J.\ Phys.\ Condens.\ Matter {\bf
1}, 901 (1989).

\end{references}
\end{document}